\documentclass[twocolumn,showpacs,preprintnumbers,amsmath,amssymb,prl, superscriptaddress]{revtex4}

\usepackage{graphicx}

\usepackage{dcolumn}
\usepackage{bm}

\begin{document}

\preprint{xxx}

\title{Structure, Scaling and Phase Transition in the Optimal Transport Network}
\author{Steffen Bohn}
\affiliation{Center for Studies in Physics and Biology,
Rockefeller University, Box 212, 1230 York Avenue, New York, NY, USA }
\affiliation{Mati\`ere et Systemes Complexes, UMR 7057
CNRS Universit\'e Paris 7 - Denis Diderot , ENS, 24 rue Lhomond,
75231 Paris Cedex 05, France} 
\author{Marcelo O. Magnasco}
\affiliation{Center for Studies in Physics and Biology,
Rockefeller University, Box 212, 1230 York Avenue, New York, NY, USA }

\date{\today}

\begin{abstract}
We minimize the dissipation rate of an electrical network under a
global constraint on the sum of powers of the conductances. We
construct the explicit scaling relation between currents and
conductances, and show equivalence to a a previous model
[J. R. Banavar {\it et al} Phys. Rev. Lett. {\bf 84}, 004745 (2000)] optimizing a power-law cost function in an abstract network. We
show the currents derive from a potential, and the scaling of the
conductances depends only locally on the currents. A numerical study
reveals that the transition in the topology of the optimal network
corresponds to a discontinuity in the slope of the power
dissipation.
\end{abstract}

\maketitle

The optimal distribution of valuables such as electricity or
telephone signals has been a subject of much study since
Westinghouse and Edison's {\sl War of the Currents} in the late
19$\rm ^th$ century \footnote{Edison, T. A., U.S.
Patent 602, 11th Feb 1880, page 4, lines 20-24 (1880), which states
"from main conductors on principal streets subsiduary main
conductors are laid through side streets; from the street
conductors, wherever desired, derived circuits are led into the
houses..."}; more recently, natural systems such as river networks
and vascular systems have been fruitfully interpreted in this light
\cite{RinaldoBook, Rinaldo1992a,Banavar2000}. Hence formal models of
optimal transport networks have attracted attention over many years
\cite{Banavar1999,Banavar2004}. However, different studies use
different definitions of network and optimize different functionals.
For example, Durand \cite{Durand2004, Durand2006} considers
hydraulic networks whose currents derive from a potential,
explicitly analogous to electrical networks; the networks are
embedded in an ambient space, and he studies the optimal geometry
and the relation between the local geometry and local topology. On
the other side, Banavar {\textit{et al.}} \cite{Banavar2000} propose
a more abstract model where the graph is not assumed to be embedded
in a target space, nor are the currents through the nodes explicitly
constrained to derive from a potential. This allows them to furnish
a strict proof that the topology of the optimized flow pattern
\cite{Banavar2000} depends on the convexity of their cost function,
but makes a direct physical interpretation of the model more
elusive. In the following, we shall introduce a third model of an
optimal transport network from whom both of these previous models
can be derived, so all formulations are, in fact, equivalent. \\
Consider an electrical transport network on a graph composed of
nodes $k$ interconnected by links $(k,l)$. There is a given current
source $i_k$ at each node and the total current input must add to
zero: $\sum_k i_k = 0$. There are variable currents $I_{kl}$ flowing
through the links; the sum of all currents impinging on a given node
$k$ must equal the given current sources: $i_k = \sum_l I_{kl}$
(Kirchhoff's current law). We associate a resistor $R_{kl } \ge 0$
to each  link $(k,l)$ and decompose its value as $R_{kl} = (d_{kl}
\kappa_{kl})^{-1}$, where $d_{kl}> 0$ is a given weight and the
conductances $\kappa_{kl}$ are  variable; considering $\kappa_{kl}$
as a conductivity per unit length, $d_{kl}$ can be thought of as the
length of the link. The dissipation rate $J$ is then a function of
the currents $I_{kl}$ through the links and the conductances
$\kappa_{kl}$:
\begin{equation}
\label{diss} J = \sum_{\textrm{links} \, (k,l)}
\frac{I_{kl}^2}{(d_{kl} \kappa_{kl})}
\end{equation}
We shall minimize this dissipation rate $J$ over the currents
$I_{kl}$ and the conductances $\kappa_{kl}$ with the local
constraint given by Kirchhoff's current law, and a supplementary
global constraint that the sum over the conductances raised to a
given power $\gamma>0$ is kept constant:
$$K^{\gamma} = \sum \kappa_{kl}^{\gamma}$$
One may interpret this constant as an amount of resources we have at
our disposal to build the network. \footnote{It
is of course possible to introduce a weight $h_{kl}$ into the
expression of the constant $K^\gamma = \sum h_{kl}
\kappa_{kl}^{\gamma}$. However, this weight can be eliminated by
straight forward rescaling of $\kappa_{kl}$ and will not change
anything in the following.} \\
\begin{figure}[b]
\begin{center}
\includegraphics[width=6cm]{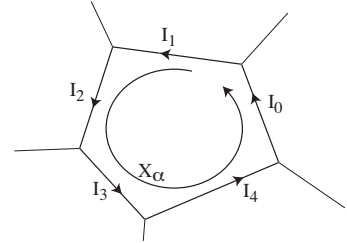}
\caption{Sketch of a loop $\alpha$ indicating the direction of the
currents. Every perturbation of the $I_{kl}$ satisfying the
constraints can be written as a weighted sum of such loops.}
\label{fig:loopAlpha}
\end{center}
\end{figure}
\begin{figure*}[t]
\begin{center}
\includegraphics[width=17cm]{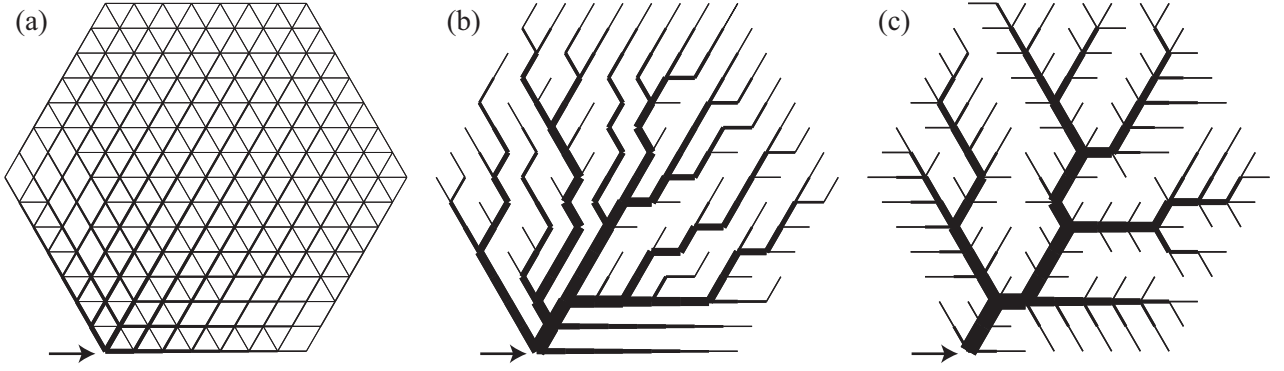}
\caption{Examples of the conductivity distributions. Results of the
relaxation algorithm with different initial conditions: (a) $\gamma
= 2.0$ and (b)  $\gamma = 0.5$. (c) Result with optimized topology
with $\gamma = 0.5$. the arrow indicated the localized inlet, the
remaining nodes are outlets with constant $i_k$. } 
\label{fig:examples}
\end{center}
\end{figure*}
Since we allow $\kappa_{kl}$ and $I_{kl}$ to vary independently, the
currents are not explicitly constrained to derive from a potential
at the nodes $U_{k}$ and Kirchhoff's voltage law (the sum of the
potential differences on a loop vanishes) need not apply. \\
Using a Lagrange multiplier $\lambda$, we define the function
$\Xi(\{\kappa_{kl}\},\{I_{kl}\})$ as
\begin{equation}
\Xi(\{\kappa_{kl}\},\{I_{kl}\}) =  \sum_{(k,l)}
\frac{I_{kl}^2}{(d_{kl} \kappa_{kl})} - \lambda \sum_{(k,l)}
{\kappa_{kl}}^{\gamma}
\end{equation}
The necessary conditions for a minima of $J$ with constant $K$ are
then:
\begin{eqnarray}
\frac{\partial \Xi}{\partial I_{kl}} = 0, \qquad \frac{\partial
\Xi}{\partial \kappa_{kl}} = 0 \label{eqn:conditions}
\end{eqnarray}
Let us first consider the derivatives with respect to $I_{kl}$. Let
$\{ \tilde{I}_{kl}\}$, $\{ \tilde{\kappa}_{kl}\}$ minimize $J$.
Adding a circular current $X_{\alpha}$ on a loop~$\alpha$ to the
currents (fig. \ref{fig:loopAlpha}) does not violate the
constraints. We (re)define the directions of the currents $
\tilde{I}_{kl}$ on the loop to be parallel to the loop current
$X_{\alpha}$. Then
\begin{equation}
0 = \frac{\partial \Xi}{\partial X_{\alpha}} \Big
\vert_{X_{\alpha}=0} =  \sum_{\textrm{loop}\, \alpha}\tilde R_{kl}
\tilde I_{kl}
\end{equation}
Thus Kirchhoff's voltage law holds at the minimum of $J$, so the
currents though the links derive from potential differences between
the nodes: $\tilde I_{kl} = \tilde R_{kl}( U_l - U_k)$. Note that
this is not the case for every arbitrary current distribution. For
instance, if all currents on the loop in fig. \ref{fig:loopAlpha}
are positive $\tilde I_{kl} > 0$, then there exists no set of
$\tilde R_{kl} \ge 0$ to fulfill this relation. \\
Let us now consider the derivatives of $\Xi$ with respect to
$\kappa_{kl}$ (eq. \ref{eqn:conditions}). With the constraint of a
constant $K$, we obtain an explicit scaling relation between the
currents and the conductivity in the minimal configuration:
\begin{equation}
\label{k_scaling} \kappa_{kl} =  \frac {(I_{kl}^2/
d_{kl})^{\frac{1}{1+\gamma}} } {\Big(\sum_{mn} {(I_{mn}^2/
d_{mn})^{\frac{\gamma}{1+\gamma}} } \Big)^{1/\gamma}} K
\end{equation}
We can now write the total dissipation (eq. \ref{diss}) in terms of
the currents alone as
\begin{equation}
\label{diss_I} {J} (\{ I_{kl}\}) = \frac{1}{K} \Big( \sum_{kl}
(I_{kl}^2/d_{kl})^{\frac{\gamma}{1+\gamma}}
\Big)^{1+\frac{1}{\gamma}}
\end{equation}
Since for $\gamma>0$, the function  $x^{1+\frac{1}{\gamma}}$ is
monotonically increasing, the original minimization problem is
reduced to the minimization of
\begin{equation}
\label{diss_I} C (\{ I_{kl}\}) = \sum_{kl}
(I_{kl}^2/d_{kl})^{\frac{\gamma}{1+\gamma}}
\end{equation}
By setting
\begin{equation}
\Gamma = \frac{2 \gamma}{\gamma + 1}
\end{equation}
and rescaling the weights as $w_{kl} =
d_{kl}^{-\frac{\gamma}{1+\gamma}}$, the quantity to be minimized is
now
\begin{equation}
\label{diss_I} C (\{ I_{kl}\}) = \sum_{kl} w_{kl}\lvert I_{kl}
\lvert^{\Gamma}
\end{equation}
which is exactly the model used by Banavar {\textit{et
al.}}\cite{Banavar2000}. They give a strict proof that for $\Gamma <
1$, the resulting structure may not have any loop, and each spanning
tree is a local minimum. For $\Gamma > 1$, there are in general
loops and a unique minimum. Due to the correspondence between
$\gamma$ and $\Gamma$, this result must apply also to our original
model where $\gamma < 1$ ($\gamma > 1$) corresponds to a $\Gamma <
1$ ($\Gamma > 1$). 
\\
On the other hand, the correspondence between the different models
allows an important conclusion about the model of Banavar {\it{et
al.}}. Since  in both formulations, the minimum is obtained by the
same set of currents, and since in our model these currents must
derive from potential differences between the nodes, this must be
true for the minimum of the Banavar {\it{et al.}} model, too. We can
furthermore write down directly the values of the corresponding
resistors as
\begin{equation}
\label{eqn:resistors} R_{kl} = (d_{kl} \kappa_{kl})^{-1} =  A \,
w_{kl} \lvert I_{kl} \rvert ^{\Gamma-2}
\end{equation}
with an arbitrary positive constant $A$. $R_{kl}$ thus scales
explicitly with the local currents for $\Gamma \ne 2$.
\\
Since positive $\gamma$ corresponds to $0 < \Gamma < 2$, the
equivalence of the two models is restricted to this parameter range.
$\Gamma > 2$ corresponds to values $\gamma < -1$, for which our
model collapses into infinitely many degenerate minima. The
relations \ref{eqn:conditions} correspond instead to a saddle node
of $J$: a minimum with respect to the $I_{kl}$ and a maximum with
respect to the $\kappa_{kl}$. Nevertheless direct inspection shows
that the current flow in the Banavar {\textit{et al.}} model is
potential with the set of resistors given by eq. \ref{eqn:resistors}
even for $\Gamma > 2$. \footnote{Given a field $\bf v$ with a curl,
a scalar field $\phi$ such that $\nabla \wedge \phi {\bf v} = {\bf
0} $ is called an integrating factor; integrating factors always
exist in two dimensions, or, in their discrete versions, for planar
graphs as in here. But while there could be in principle a set of
resistors that would make arbitrary currents in a planar model
derive from a potential, such resistors would neither be guaranteed
to be positive nor to depend only locally on the currents, as our
result shows.} 
\\
In order to get a deeper insight into the transition at $\gamma =
1$, we search numerically for the minimal dissipation configuration
of an example network, a triangular network of conductivities with a
hexagonal border, with equal weights $d_{kl} \equiv 1$.  The total
number of nodes $N_{nodes}$ scales roughly as the square of the
linear dimension of the network, given by the diameter of the graph
$N_{dia}$. Except for those on the border, each node is linked
by conductivities to six neighboring nodes.
\\
We place a current source at a corner of the hexagon ($i_0$), the
remaining $(N_{nodes}-1)$ nodes present homogeneous distributed
sinks; each node absorbs $i_k = - i_0/(N_{nodes}-1)$.
\\
As an order parameter, we will consider the normalized dissipation
rate $J_{min}/J_{homo}$, where $J_{homo}$ is the total dissipation
with a constant conductivity distribution $\kappa_{kl} \equiv
\textrm{const.}$, and  $J_{min}$ is the dissipation for the
optimized distribution of the conductivities. Note that $J_{homo}$
corresponds also to $\gamma \rightarrow \infty$. \\
\begin{figure}
\begin{center}
\includegraphics[width=7.5cm]{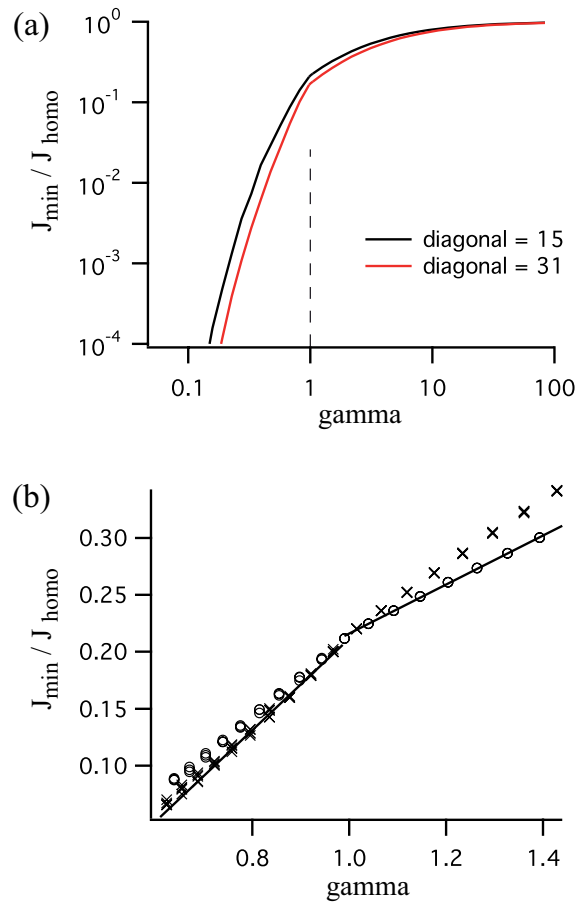}
\caption{ (a) The normalized minimum dissipation rate
$J_{min}/J_{homo}$ as a function of $\gamma$ for a network with
$N_{dia} = 15$ (462 links) and a network with $N_{dia} =
31$ (2070 links, in red). Note the discontinuity of the slopes at
$\gamma = 1$. (b) A detailed view of the crossover at $\gamma = 1$
for $N_{dia}=15$. Cross symbols show data points obtained by
optimizing a tree topology, circles show the output of the
relaxation algorithm. The continuous lines indicate the actual
minimum.} \label{fig:diss2D}
\end{center}
\end{figure}
The previous discussion allows us to simplify the minimization
problem enormously: using the scaling relation between $\kappa_{kl}$
and $I_{kl}$, one can restrict the search of the minimum to the
space of the currents or the space of conductivities. Furthermore,
we can use the fact that the optimized current distribution derives
from a potential $U_k$ to construct a simple relaxation algorithm.
Starting with a random distribution of $\kappa_{kl}$, we calculate
first the values of the potential at the nodes by solving the system
of linear equations $i_k = \sum_{l} R_{kl} (U_k - U_l)$, then the
currents through the  links $I_{kl}$ are determined. We use these
currents to determine a first approximation of the optimal
conductivities on the basis of the scaling relation. Then, the
currents are recalculated with this set of conductivities, and the
scaling relation is reused for the next approximation. These steps
are repeated until the the values have converged.  We check by
perturbing the solution that it actually is a minimum of the
dissipation, which was always the case. \\
For  all $\gamma>1$, independently of the initial conditions, the
same conductivity distribution is obtained, which conforms to the
analytical result of \cite{Banavar2000}: there exists a unique
minimum which is therefore global. \\
Furthermore, the distribution of $\kappa_{kl}$ is ``smooth'',
varying only on a ``macroscopic scale'', as show in
Fig.~\ref{fig:examples}(a)). No formation of any particular
structure occurs. However,  the conductivity distribution is not
isotropic. We can interpret the conductivity distribution as a
discrete approximation of a continuous, macroscopic conductivity
tensor (see also \cite{Durand2004}). The smooth aspect of the
distribution is conserved while approaching $\gamma \rightarrow 1$
while the local anisotropy increases, while the values of all
$\kappa_{kl}$ remain finite, even if they get very small. For
$\gamma = 1.5$ and $N_{dia} = 15$,  the conductivity
distribution spreads already over eight decades and becomes still
broader as $\gamma\rightarrow 1^+$, in which limit the number of
iteration steps diverges as the minima becomes less and less steep. \\
$\gamma = 1$ presents a marginal case. The results of the simulation
suggest that the minimum is highly degenerate, i.e., there are a
large number of conductivity distributions yielding the same minimal
dissipation. \\
For  $\gamma<1$, the output of the  relaxation algorithm is qualitatively
different (fig.~\ref{fig:examples}(b)). The currents are canalized
in a hierarchical manner: a large number of conductivities rapidly
converge to zero and thus vanish transforming the topology from a
highly redundant network to a spanning tree. This, too, is predicted
by the analytical results \cite{Banavar2000}. In contrast to $\gamma
> 1$, the conductivity distribution can not be interpreted as a
discrete approximation of a conductivity tensor: for $N_{dia}
\rightarrow \infty$, the structure becomes fractal. \\
For different initial conditions, the relaxation algorithm yields
trees with different topologies: each local minima in the
high-dimensional and continuous space of conductivities $\{\kappa_{kl}\}$
correspond to a different tree topology. Given a tree topology, the
currents through the links are given directly by the topology and do
not depend on the values of the $\kappa_{kl}$, and so using the
scaling relation, one may directly write down the dissipation rate
for a given tree. For $\gamma < 1$, we do thus not need to apply the
relaxation algorithm, but we should search for the global minima in
the (exponentially large) space of tree topologies using a
Monte-Carlo algorithm. This regime has been widely explored in the
context of river networks \cite{Rinaldo1992a, Sun1994, RinaldoBook},
mainly for a set of parameters that corresponds, in our case, to
$\gamma = 0.5$. An example of a resulting minimal dissipation tree
structure is given in fig.~\ref{fig:examples}(c). Note also, that
the scaling relations can be seen as some kind of erosion model: the
more currents flows trough a link, the better the link conducts
\cite{RinaldoBook}. \\
The qualitative transition is reflected also quantitatively in the
value of the minimal dissipation (fig~\ref{fig:diss2D}(a)). The
points for $\gamma >1$ were obtained with the relaxation algorithm,
the points $\gamma <1$ by optimizing the tree topologies with a
Monte-Carlo algorithm. For $\gamma \rightarrow \infty$,
$J_{min}/J_{homo} \rightarrow 1$ by definition; for $\gamma
\rightarrow 0$,  $J_{min}/J_{homo} \rightarrow 0$, because the
vanishing $\kappa_{kl}$ allow the the remaining $\kappa_{kl}
\rightarrow \infty$. \\
Figure~\ref{fig:diss2D}(b) shows the behavior of minimal
dissipation rate close to $\gamma = 1$. For $\gamma$ smaller than
one, the relaxation method spends a long time only to furnish a
local minimum, while the Monte-Carlo algorithm searching for the
optimal tree topologies gives lower dissipation values. The
different values corresponding to different realization indicate
that the employed Monte-Carlo method does not find the exact global
minima. For $\gamma>1$, the relaxation algorithm gives the lower
$J_{min}$ because the global minima does not have a tree topology. \\
While the curve is continuous, the crossover at $\gamma = 1$ shows a
clear change in the slope of $J_{min}(\gamma)$. One could interpret
this behavior as a second order phase transition. (The change in
slope is of course preserved in the function $C({I_{kl}})$ used by
\cite{Banavar2000}.) \\
As an intriguing practical application of these models, one may for
instance cite the venation of plant leaves. Experimental evidence
\cite{Zwieniecki2002} shows that the water transport through the
veins derives from a pressure gradient. The venation pattern however
shows a enormous redundancy of loops \cite{esaubook,
RothNebelsick2001,Couder2002}. On the basis of some examples, it has
been proposed \cite{Roth1995,RothNebelsick2001} that the loops are
actually meaningful to optimize the water transport in the leaf. The
results presented in this paper however shows that this is not the
case: optimization either leads to a tree topology, or to no
structure at all. If the venation pattern is really based a
optimization principe, it cannot simply be optimization of a steady
state water transport, even if Murray's law seems to hold at the
nodes of the venation \cite{McCulloh2003}.

\bibliography{flow}

\end{document}